\documentclass[%
 reprint,
 amsmath,amssymb,
 aps,
pra,
]{revtex4-2}

\usepackage{graphicx}% Include figure files %HN to use Japanese comments
\usepackage{dcolumn}% Align table columns on decimal point
\usepackage{bm}% bold math
\usepackage{color}

\newcommand{\YCOM}[1]{{{\color{red}[#1] }}}
\renewcommand{\YCOM}[1]{}

\renewcommand{\i}{{\mathrm{i}}}

\begin{document}

\preprint{APS/123-QED}

\title{
Global Phase Synchronization Decoupled from Amplitude Dynamics
}

\author{Koichiro Yawata}
\affiliation{Department of Systems and Control Engineering and Research Center for Autonomous Systems Materialogy, Institute of Science Tokyo, Tokyo 152-8552, Japan}

\author{Hiroya Nakao}
\affiliation{Department of Systems and Control Engineering and Research Center for Autonomous Systems Materialogy, Institute of Science Tokyo, Tokyo 152-8552, Japan}

\begin{abstract}
The Kuramoto model is a canonical framework for analyzing phase synchronization, yet its utility is restricted to the vicinity of the oscillator's unperturbed limit cycle. Here, we present a method to construct coupled-oscillator models that globally preserve Kuramoto-type phase dynamics by using phase-amplitude coordinates defined via Koopman operator theory.
We introduce a solvable model, termed Kuramoto-Stuart-Landau model, which exhibits nontrivial synchronized dynamics far from the limit cycle.
We also construct three phase synchronized oscillators whose amplitudes exhibit Lorenz-type chaos.
Our method is applicable to general limit-cycle systems, achieving phase synchronization globally while preserving arbitrarily complex amplitude dynamics.
\end{abstract}

\maketitle

\textbf{Introduction.}
Rhythmic phenomena and their synchronization are widely observed in diverse natural and engineered systems, including neuronal and cardiac activities, circadian clocks, chemical oscillations, and power grids~\cite{Winfree2001geometry,Kuramoto1984chemical,Pikovsky2001synchronization,strogatz2004sync,Kralemann2013,stankovski2015coupling,pecora1990synchronization,rosenblum1996phase,boccaletti2002synchronization,dorfler2013synchronization,motter2013spontaneous,womelsdorf2007modulation}.
These systems are typically modeled as nonlinear oscillators 
that exhibit stable limit cycles.
To analyze the synchronization dynamics of nonlinear oscillator systems, phase reduction theory provides a simple and powerful mathematical method~\cite{Winfree2001geometry,Kuramoto1984chemical,Nakao2016phase,Ermentrout2010mathematical,monga2019phase,yawata2024phase}.
This method reduces the dimensionality of the oscillator by projecting the dynamics onto a single phase variable in the vicinity of the limit cycle.
Based on the reduced phase equation, the effects of weak coupling on the oscillators can be systematically analyzed, providing a theoretical foundation for understanding and controlling synchronization 
dynamics~\cite{Monga2019optimal,kiss2007engineering,takata2021fast,dasanayake2011optimal,yawata2025data}.

The Kuramoto-type phase model is the most widely used model for studying synchronization of coupled oscillators derived via phase reduction~\cite{Kuramoto1984chemical,acebron2005kuramoto}.
Despite its simplicity, it captures essential features of synchronization dynamics, including the collective synchronization transition of many oscillators.
However, as a phase-based model, its applicability is restricted to the vicinity of the underlying limit cycle; it cannot describe the oscillator dynamics away from the limit cycle and necessitates a weak coupling assumption between the oscillators.
Indeed, the significance of amplitude dynamics in coupled limit-cycle oscillators has been 
investigated~\cite{aronson1990amplitude,koseska2013oscillation,matthews1990phase}. 
Also, in the context of coupled chaotic oscillators, phase synchronization with complex amplitude dynamics can be observed~\cite{rosenblum1996phase,Pikovsky2001synchronization,boccaletti2002synchronization}.

To gain further insight into synchronization away from the limit cycle, phase-amplitude reduction has attracted considerable attention recently~\cite{mauroy2018global,shirasaka2017phase,wilson2016isostable,wilson2018greater,wilson2020phase}.
Phase-amplitude reduction is based on Koopman operator theory and introduces not only the phase but also the amplitude coordinates, which can describe deviations of the oscillator state from the limit cycle.
However, to derive a closed set of phase-amplitude equations~\cite{shirasaka2017phase,wilson2016isostable,wilson2018greater,wilson2020phase}, the assumption of weak coupling is still necessary, limiting its applicability to the vicinity of the limit cycle.

In this study, we propose a framework for constructing coupled-oscillator models that globally preserve Kuramoto-type phase dynamics over the entire basin of attraction of the limit cycle, using phase-amplitude coordinates.
We first show how to design coupling functions that exactly yield the Kuramoto phase dynamics  for arbitrary oscillators irrespective of the amplitude dynamics.
We then introduce an analytically tractable example, the Kuramoto--Stuart--Landau (KSL) model, and analyze both the two-oscillator and many-oscillator settings under strong coupling, revealing nontrivial synchronized dynamics away from the limit cycle.
Finally, we present a three-oscillator example with Kuramoto phase dynamics and Lorenz-type chaotic amplitude dynamics.\\
\par

\textbf{Method.}
We consider a limit-cycle oscillator described by the following differential equation,
\begin{align}
    \dot{\bm{x}}=\bm{F}(\bm{x}),
    \label{eq:LC}
\end{align}
where $\bm{x}(t)\in \mathbb{R}^D$ represents the state of the oscillator at time $t$ in a $D$-dimensional state space and $\bm{F}(\bm{x}):\mathbb{R}^D\to\mathbb{R}^D$ is a smooth vector field representing the oscillator dynamics.
We assume that this dynamical system has an exponentially stable limit-cycle solution $\bm{x}_0(t)$ and denote its basin of attraction by $B\subseteq\mathbb{R}^D$.

In the Koopman operator framework~\cite{mauroy2018global,shirasaka2017phase,wilson2016isostable,koopman1931hamiltonian,mezic2005spectral,brunton2021modern}, general observables of nonlinear dynamical systems can be decomposed using Koopman eigenfunctions, yielding linearly evolving Koopman modes.
For an exponentially stable limit cycle, the principal Koopman eigenvalues can be taken as $\lambda_0 = i \omega$ and $\lambda_1,\ ...,\ \lambda_{D-1}$, where $\omega$ is the natural frequency and $\lambda_{1, ..., D-1}$ are the Floquet exponents with negative real parts,
respectively~\cite{mauroy2018global}. 
For simplicity, we assume $\lambda_{1, ..., D-1}$ to be real and distinct. Using the corresponding Koopman eigenfunctions, we can introduce an asymptotic phase function $\Theta(\bm{x}):B\to[0,2\pi)$ and $D-1$ amplitude functions $\Gamma^{(k)}(\bm{x}):B\to \mathbb{R}$
in the basin $B$ of the limit cycle~\cite{mauroy2018global}, satisfying
\begin{align}
\bm{F}(\bm{x}) \cdot \nabla {\Theta}(\bm{x}) = \omega,
\quad
\bm{F}(\bm{x}) \cdot \nabla {\Gamma}^{(k)}(\bm{x}) = \lambda^{(k)} \Gamma^{(k)}(\bm{x})
\label{Eq:Koopman}
\end{align}
for $k=1, ..., D-1$, where $\nabla = \partial / \partial \bm{x}$ represents the vector gradient and $\bm{F}(\bm{x}) \cdot \nabla$ is the infinitesimal Koopman operator of Eq.~\eqref{eq:LC}. 
We can then define the asymptotic phase~\cite{Winfree2001geometry,Kuramoto1984chemical,Nakao2016phase,Ermentrout2010mathematical} and amplitude coordinates~\cite{mauroy2016global,mauroy2018global,shirasaka2017phase,wilson2016isostable} of the oscillator state $\bm{x}$ as
\begin{align}
    &\theta = \Theta(\bm{x}),
    \quad
    r^{(k)} =\Gamma^{(k)}(\bm{x}).
\end{align}
It is then clear that $\dot{\theta} =  {\bm F}({\bm x}) \cdot \nabla \Theta({\bm x}) = \omega$ and $\dot{r}^{(k)} = {\bm F}({\bm x}) \cdot \nabla \Gamma^{(k)}({\bm x})  = \lambda^{(k)} r^{(k)}$ for any $\bm{x} \in B$.
The amplitudes $r^{(k)}$ quantify the departure of the oscillator state from the limit cycle into $D-1$ different directions and vanish when the oscillator state is on the limit cycle.
Thus, these phase and amplitude coordinates 
%gives
provide a globally linearized description of the oscillator. 

We introduce gradient vectors of the phase and amplitude functions evaluated at the oscillator state $\bm{x}$ as
\begin{align}
\label{eq:sensitivity}
\bm{Z} (\bm{x})
=\nabla \Theta(\bm{x}),
\quad
\bm{I}^{(k)}
{(\bm{x})}
{=\nabla \Gamma^{(k)} {(\bm{x})}}.
\end{align}
In the phase-amplitude reduction theory for weakly perturbed oscillators~\cite{wilson2016isostable,shirasaka2017phase,mauroy2013isostables,mauroy2016global,mauroy2018global,lan2013linearization}, these gradient vectors evaluated on the limit cycle at $\bm{x} = \bm{x}_0(\theta/\omega)$ are of particular importance and called the phase and amplitude sensitivity functions.
Here, we emphasize that they are defined for a general state $\bm{x} \in B$.
We can assume that, at each ${\bm x} \in B$, the vectors ${\bm Z}(\bm{x}), {\bm I}_1(\bm{x}), ..., {\bm I}_{D-1}(\bm{x})$ are linearly independent from the properties of the phase-amplitude coordinates~\cite{mauroy2016global,wilson2020phase}.
Therefore, we can define the dual vectors $\bm{u}{(\bm{x})} \in \mathbb{R}^D$ and $\bm{v}^{(k)}{(\bm{x})} \in \mathbb{R}^D$ ($k=1, ..., D-1$) of $\bm{Z}(\bm{x})$ and $\bm{I}^{(k)}(\bm{x})$ satisfying
\begin{align}
    &\bm{u}(\bm{x}) \cdot \bm{Z}(\bm{x}) = 1,\quad
    \bm{u}(\bm{x}) \cdot \bm{I}^{(k)}(\bm{x}) = 0,\cr
    &\bm{v}^{(k)}(\bm{x}) \cdot \bm{Z}(\bm{x}) = 0,\quad
    \bm{v}^{(k)}(\bm{x}) \cdot \bm{I}^{(l)}(\bm{x}) = \delta_{kl},
\label{eq:dual}
\end{align}
for $\bm{x} \in B$, where $\delta_{kl}$ is Kronecker's delta.
The vector $\bm{u}(\bm{x})$ represents the direction at $\bm{x}$ in which the phase increases by one per unit time without altering the amplitude, i.e., 
$(d/dt) \Theta( \bm{x} + \bm{u}(\bm{x})  t ) = \nabla \Theta(\bm{x}) \cdot {\bm{u}}(\bm{x}) = \bm{Z}(\bm{x}) \cdot \bm{u}(\bm{x}) = 1$.
From Eq.~\eqref{Eq:Koopman}, this direction is parallel to $\bm{F}(\bm{x})$, i.e., tangent to the orbit of the oscillator state.
Similarly, the vector $\bm{v}^{(k)}$ represents the direction in which the $k$th amplitude increases by one per unit time without affecting the phase and other amplitudes.

Let us consider a system of $N$ coupled limit-cycle oscillators described by
\begin{align}
    \dot{\bm{x}}_{i} = \bm{F}_{i}(\bm{x}_{i}) + \bm{H}_i ({\bm x}_{1},...,{\bm x}_{N})
    \label{eq:coupledosc}
\end{align}
for $i=1, ..., N$. Here, $\bm{x}_{i}$ is the $i$th oscillator state,
$\bm{F}_{i}$ is the dynamics of $i$th oscillator,
and $\bm{H}_{i}( {\bm x}_{1},...,{\bm x}_{N} )$ is a coupling function 
representing the effect of the other oscillators state on $\bm{x}_{i}$
.
Note that $\bm{F}_{i}$ and consequently the limit cycle, natural frequency, Floquet exponents, phase and amplitudes, and gradient and dual vectors, also depend on $i$; they are denoted by the subscript $i$.

Our aim is to find the function $\bm{H}_{i}$ that always realizes the phase dynamics of the Kuramoto type, even when each oscillator state is away from the limit cycle and the amplitudes $r^{(k)}$ are non-zero.
In this study, we choose the coupling function as
\begin{align}
\label{eq:coupling0}
    \bm{H}_{i}({\bm x}_{1},...,{\bm x}_{N}) 
%    = K_{ij} \sin ( \Theta_{j}(\bm{x}_{j}) - \Theta_{i}( \bm{x}_{i} ) ) \bm{u}_{i}(\bm{x}_{i}),
    =&\sum_{j=1}^N K_{ij} \sin ( \Theta_{j}(\bm{x}_{j}) - \Theta_{i}( \bm{x}_{i} ) ) \bm{u}_{i}(\bm{x}_{i}) \cr
&\quad+
\sum_{\ell=1}^{D-1}  G_i^{(\ell)}(\bm{x}_{1}, ..., \bm{x}_{N}) \bm{v}^{(\ell)}_{i}( \bm{x}_{i} ),
\end{align}
where $K_{ij} \in \mathbb R$ is the coupling strength and $G_i^{(\ell)} : {\mathbb R}^{D \times N} \to {\mathbb R}$ is an arbitrary function of the oscillator states.
Then, the corresponding phase equation is exactly of the Kuramoto type. Indeed, 
by using the biorthogonality relations in Eq.~\eqref{eq:dual},
the phase $\theta_{i} = \Theta_{i}( \bm{x}_{i} )$ of the $i$-th oscillator obeys
\begin{align}
 \dot{\theta}_{i} 
&= 
\nabla \Theta_{i}( \bm{x}_{i} ) \cdot \Big( \bm{F}_{i}(\bm{x}_{i}) + \bm{H}_{i}({\bm x}_{1},...,{\bm x}_{N}) \Big)
\cr
&= \omega + \sum_{j=1}^N  \bm{Z}_{i}(\bm{x}_{i}) \cdot \bm{u}_{i}(\bm{x}_{i}) K_{ij}\sin ( \Theta_{j}(\bm{x}_{j}) - \Theta_{i}( \bm{x}_{i} ) )  
\cr
&= \omega + \sum_{j=1}^N K_{ij} \sin ( \theta_{j} - \theta_{i} ).
\end{align}
Similarly, the equations for the $k$th amplitudes of the $i$-th oscillator
${r}^{(k)}_{i} = \Gamma^{(k)}_{i}(\bm{x}_{i})$ is given by
\begin{align}
\dot{r}^{(k)}_{i} 
&=  \nabla \Gamma_i^{(k)}( \bm{x}_{i} ) \cdot \Big( \bm{F}_{i} + \bm{H}_{i}({\bm x}_{1},...,{\bm x}_{N}) \Big)
\cr
&= \lambda^{(k)}_{i} r^{(k)}_{i} +  \sum_{\ell=1}^{D-1} \bm{I}^{(k)}_i(\bm{x}_{i} ) \cdot\bm{v}^{(\ell)}_{i}( \bm{x}_{i} ) G_i^{(\ell)}(\bm{x}_{1}, ..., \bm{x}_{N}) 
\cr
&= \lambda^{(k)}_{i} r^{(k)}_{i} + \sum_{j=1}^N G_i^{(k)}(\bm{x}_{1}, ..., \bm{x}_{N}) 
%\quad
%(k=1, ..., D-1).
\end{align}
for $k=1, ..., D-1$.
Here, the function $G^{(k)}_{i}$ determining the amplitude dynamics is not yet specified, but it does not affect the phase dynamics.
Note that these equations are exact and valid in the strong coupling regime.

Thus, by using the coupling function of Eq.~\eqref{eq:coupling0}, the coupled-oscillator system Eq.~\eqref{eq:coupledosc} always obeys the Kuramoto phase dynamics irrespective of their amplitudes in the whole basin of the limit cycle of each oscillator.
This contrasts with the conventional Kuramoto model, which is typically derived via phase reduction of weakly coupled limit-cycle oscillators and holds only near the unperturbed limit cycle.\\
\par

\textbf{Kuramoto-Stuart-Landau Model.}
As an analytically tractable example, we consider a system of 
Stuart-Landau (SL) oscillators with the proposed coupling function. This model represents the normal form of dynamical systems near a supercritical Hopf bifurcation and is widely used as a canonical model of limit-cycle oscillators~\cite{Kuramoto1984chemical,Nakao2016phase,kato2021asymptotic}.
Each oscillator state $\bm{x} = (x, y)$ obeys
\begin{align}
\dot{x} &= a x - b y - (c x - d y)(x^2 + y^2), \\
\dot{y} &= b x + a y - (d x + c y)(x^2 + y^2),
\end{align}
where $a>0$, $b$, $c>0$, and $d$ are real parameters.
The limit cycle is a circle of radius $\sqrt{a/c}$ centered at the origin, $x_0^2 + y_0^2 = a/c$, whose natural frequency and Floquet exponent are $\omega = b -ad/c$ and $\lambda=-2a$, respectively. 

The asymptotic phase and amplitude functions are explicitly given by~\cite{kato2021asymptotic} 
\begin{align}
\Theta({\bm x}) &= \arctan\left( \frac{y}{x} \right) - \frac{d}{2c}\log \left( \frac{c}{a}(x^2 + y^2) \right)
\label{eq:phasefunction},
\\
\Gamma({\bm x}) &=  c - \frac{a}{x^2+y^2}\label{eq:ampfunction},
\end{align}
where we have chosen ${\bm x} = (\sqrt{a/c}, 0)$ as the phase origin, i.e., $\Theta(\sqrt{a/c}, 0) = 0$. The amplitude vanishes on the limit cycle, i.e., $\Gamma(x_0, y_0) = 0$.
Note that only a single amplitude exists because $D=2$; hence, we omit the superscript $(k)$ from the amplitude $r = \Gamma({\bm x})$ and other quantities in what follows.
In the absence of coupling, the phase and amplitude simply obey
\begin{align}
    \theta(t) = \theta(0) + \omega t,
    \quad
    r(t) = e^{\lambda t} r(0).
\end{align}
From the phase $\theta$ and amplitude $r$, it is also possible to convert back to $x$ and $y$ as
\begin{equation}\label{eq:reverse2xy}
\begin{aligned}
	x &= \sqrt{ \frac{a}{c - r} } \cos\left( \theta + \frac{d}{2c} \log\left( \frac{c}{c - r} \right) \right), \\
	y &= \sqrt{ \frac{a}{c - r} } \sin\left( \theta + \frac{d}{2c} \log\left( \frac{c}{c - r} \right) \right).
\end{aligned}
\end{equation}

The gradient vectors of $\Theta({\bm x})$ and $\Gamma({\bm x})$ are given by
\begin{align}
\bm{Z}({\bm x}) = \frac{1}{c(x^2+y^2)} \begin{bmatrix} - d x - c y \\ c x - d y \end{bmatrix},
\quad
\bm{I}({\bm x}) = \frac{2a}{(x^2+y^2)^2} \begin{bmatrix} x \\ y \end{bmatrix},
\end{align}
and the dual vectors can be taken as
\begin{align}\label{eq:g}
\bm{u}({\bm x}) = 
\begin{bmatrix}
-y \\
x
\end{bmatrix},
\quad
{\bm{v}}({\bm x}) = \frac{x^2+y^2}{2{ac}}
\begin{bmatrix}
{cx-dy} \\
{dx+cy}
\end{bmatrix}.
\end{align}
Note that these dual vectors vanish at the origin $(x, y) = (0, 0)$, where the phase and amplitude are undefined and the oscillator state is outside the basin $B$.
\par

In the absence of amplitude dynamics, the coupling function $\bm{H}_{i}$ in Eq.~\eqref{eq:coupling0} yielding the Kuramoto dynamics is explicitly given by
\begin{align}\label{eq:coupling}
    \bm{H}_{i}&(\bm{x}_{1},...,\bm{x}_{N})
    = \sum_j^NK \sin (\Theta_{j}({\bm x}_{j})-\Theta_{i}({\bm x}_{i})) \bm{u}(x_{i},y_{i}) \cr
    &= \sum_j^NK\Big(\frac{x_{i} y_{j} - x_{j} y_{i}}{\rho_{i}\rho_{j}} \cos\gamma_{i}
    + \frac{x_{j} x_{i} + y_{j} y_{i}}{\rho_{i}\rho_{j}} \sin\gamma_{i}
    \Big)
    \begin{bmatrix}
-y_{i} \\
x_{i}
\end{bmatrix},
\end{align}
where $\rho_{i}=\sqrt{x_{i}^2+y_{i}^2}$ and $\gamma_{i} = ( d_{i} / c_{i} ) \log ( \rho_{i} / \rho_{j} )$.
Here, the quantity $\gamma_{i}$ characterizes the amplitude ratio of oscillators $i$ and $j$, compensating for the asymptotic phase shift induced by the amplitude difference.

If $c_{i}$ and $d_{i}$ are identical for all oscillators, the functional form of the coupling function $\bm{H}_{i}$ takes the same functional form for all $i$ and $j$.
In what follows, we set $a$, $c$ and $d$ to be identical across oscillators and assume $a = c$. Then, each oscillator's limit cycle is always on the unit circle, and its natural frequency $\omega_{i}$ can be varied by appropriately adjusting the parameter $b_{i}$ without affecting the limit cycle and the coupling function.\\
\par

\textbf{Two-coupled KSL oscillators.}
First, we consider a pair of KSL oscillators with identical properties,
\begin{align}
\dot{\bm{x}}_1 &= \bm{F}(\bm{x}_{1}) + \bm{H}_{1}(\bm{x}_{1}, \bm{x}_{2}),
\cr
\dot{\bm{x}}_2 &= \bm{F}(\bm{x}_{2}) + \bm{H}_{2}(\bm{x}_{1}, \bm{x}_{2}),
\end{align}
where $\bm{x}_{i} = (x_{i}, y_{i})$ for $i=1, 2$,  $\bm{H}_{1}(\bm{x}_{1}, \bm{x}_{2}) = K \sin( \Theta(\bm{x}_{2}) - \Theta(\bm{x}_{1}) ) \bm{u}(\bm{x}_{1})$, and
$\bm{H}_{2}(\bm{x}_{1}, \bm{x}_{2}) = K \sin( \Theta(\bm{x}_{1}) - \Theta(\bm{x}_{2}) ) \bm{u}(\bm{x}_{2})$.
The corresponding phase equations for $\theta_1 = \Theta(\bm{x}_{1})$ and $\theta_2 = \Theta(\bm{x}_{2})$ are then
\begin{align}
\dot{\theta}_1 = \omega + K \sin(\theta_2 - \theta_1), 
\cr
\dot{\theta}_2 = \omega + K \sin(\theta_1 - \theta_2),
\end{align}
where $\omega=b-ad/c$.
The mean phase $\phi=(\theta_1+\theta_2)/2$ and phase difference $\psi=\theta_1-\theta_2$ obey
$
    \dot{\phi} = \omega,
    \quad
    \dot{\psi}=-2K\sin{\psi},
$
which can be solved as
\begin{align}
\phi(t) &= \phi(0)+\omega t,\\
\psi(t) &= 2 \arctan\left( \tan\frac{\psi(0)}{2} \, e^{-2Kt} \right).
\end{align}
The amplitude equations for $r^{(1,2)} = \Gamma(\bm{x}^{(1,2)})$ are simply $\dot{r}_1 = \lambda {r}_1$ and $\dot{r}_2 = \lambda {r}_2$ where $\lambda = -2a$.
Thus, the phase-amplitude dynamics of the two oscillators are
\begin{align}
\theta_1(t) = \phi(t)-\psi(t)/2,
\quad
r_1(t) = r_1(0)e^{\lambda t},
\cr
\theta_2(t) = \phi(t)+\psi(t)/2,
\quad
r_2(t) = r_2(0)e^{\lambda t}.
\end{align}

\begin{figure}
    \centering
    \includegraphics[width=0.8\linewidth]{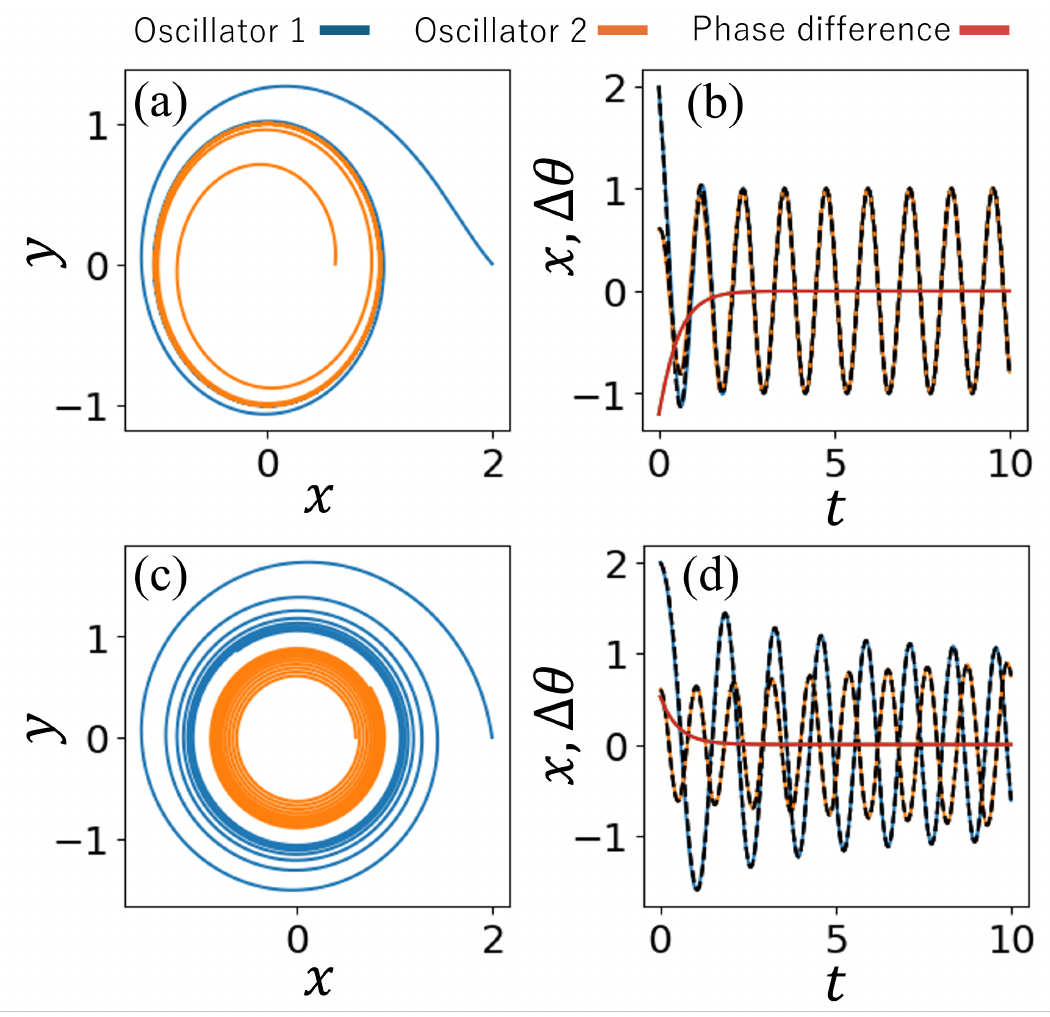}
    \caption{Dynamics of a pair of coupled KSL oscillators. (a,c): Trajectories in the $(x, y)$ plane. (b,d): Temporal evolution of $x$ and the phase difference $\Delta \theta$.  The parameter $a$ is set to $1.0$ in (a, b) and $0.1$ in (c, d).
    Blue and orange lines indicate numerical results for oscillators~1 and~2, respectively. The black dashed lines represent the analytical results, and the red line shows the phase difference.
}
    \label{fig:two-body}
\end{figure}

Figure~\ref{fig:two-body} illustrates the effect of varying $a$ (with $a = c$) on the dynamics of coupled KSL oscillators.
The other parameters are set to $K=1, b=2\pi$ and $d=1$.
In panels (a) and (b) for $a=1$, the relatively large decay rate
($\lambda = -2$) 
rapidly drives the oscillator states onto the limit cycle, where phase synchronization is then established, leading to complete synchronization.
This represents the standard scenario of phase synchronization for weakly interacting oscillators,
where the coupling is weak relative to the amplitude decay rate.
In contrast, in panels (c) and (d) for $a=0.1$, the decay rate is much smaller ($\lambda = -0.2$). This yields an intriguing behavior: although the temporal waveforms of the two oscillators, started from inside and outside of the limit cycle, appear to be non-synchronous due to large amplitude differences, the two oscillators are actually synchronized with respect to their asymptotic phases. This is confirmed by the convergence of $\Delta \theta = \theta_1 - \theta_2$ to $0$. After a sufficiently long transient, the two oscillator states eventually converge to the limit cycle, and complete synchronization is observed.\\ 
\par

\textbf{Population of KSL oscillators.}
We now consider a system of $N$ coupled KSL oscillators described by
\begin{align}
\dot{\bm{x}}_{i} = \bm{F}_{i}(\bm{x}_{i}) + \bm{H}_{i}(\bm{x}_{1},...,\bm{x}_{N}),
\end{align}
for $i=1, ..., N$ and choose $\bm{H}_{i}(\bm{x}_{1},...,\bm{x}_{N}) = (K/N)\sum_{j=1}^N\sin( \Theta(\bm{x}_{j}) - \Theta(\bm{x}_{i}) ) {\bm{u}_{i}}(\bm{x})$, where $K$ is the global coupling strength.
Then, the phase dynamics is exactly given by the classical Kuramoto model,
\begin{align}
\dot{\theta}_{i} = \omega_{i} +  \frac{K}{N} \sum_{j=1}^N \sin(\theta_{j} - \theta_{i}).
\label{eq:kuramoto}
\end{align}
%
%i.e., the Kuramoto model. 
Note that $\bm{F}_{i}$ and consequently the limit cycle and other quantities explicitly depend on the oscillator index $i$.
We assume that each $\omega_{i}$ is randomly and independently drawn from the Lorentzian distribution,
%probability density function (PDF),
%
\begin{align}
    \label{eq:omega_distribution}
    g(\omega) = \frac{\Delta}{\pi} \frac{1}{(\omega-\omega_0)^2+\Delta^2},
\end{align}
where $\omega_0$ is the central frequency and $\Delta$ is the half-width at half-maximum.
In the original KSL model, the parameter $b_{i}$ is adjusted to realize the given frequency $\omega_{i}$.

We employ the Ott-Antonsen ansatz~\cite{ott2008low} to analyze this model. 
The collective synchronization of this system is quantified by the complex order parameter $\zeta$, or real order parameters $R$ and $\Psi$, given by
\begin{align}
    \zeta=Re^{{\i}\Psi}=\frac{1}{N}\sum_{i}^N\exp({\i}\theta_{i}),
\end{align}
where {$\i=\sqrt{-1}$} is the imaginary unit.
Assuming that the initial probability density function (PDF) of the oscillator phases lies on the Poisson manifold~\cite{ott2008low}, the complex order parameter $\zeta$ obeys
\begin{align}
    \dot{\zeta} = ({\i}\omega_0-\Delta)\zeta + \frac{K}{2}(\zeta-|\zeta|^2\zeta),
\end{align}
and {correspondingly $R$ and $\Psi$ obey}
\begin{align}
    \dot{R} = (-\Delta+\frac{K}{2})R-\frac{K}{2}R^3,
    \quad
    \dot{\Psi}=\omega_0.
\end{align}
When $K > 2 \Delta$, there exists a stable fixed point $R={R_c}=\sqrt{1-2\Delta/K}$ corresponding to the steady collectively synchronized state.

To demonstrate the nontrivial synchronization behavior of this model, we initialize the oscillator states such that $R = {R_c}$. Specifically, we sample the initial phases of the oscillators from the following PDF:
\begin{align}\label{eq:theta_dis}
    f(\theta)=\frac{1-R_c^2}{2\pi(1-2R_c\cos(\theta-\Psi(0))-R_c^2},
\end{align}
where $\Psi(0)$ is the initial value of $\Psi$. Then, the subsequent evolution of the order parameters are simply $R(t)={R_c}$ and $\Psi(t)=\Psi(0)+\omega_0t$.
Rewriting Eq.~(\ref{eq:kuramoto}) with the order parameter gives
\begin{align}
    \label{eq:periodic_force}
    \dot{\theta_{i}} = \omega_{i} + K R_c \sin (\Psi(t) - \theta_{i}),
\end{align}
which can be interpreted as a single oscillator driven by a periodic external force.
Both phase-locked and drifting solutions of the above equation can be analytically obtained. The drifting solution is given by
\begin{align}
    \theta_{i} &= \omega_0 t + 2\tan^{-1}\left( \frac{K'}{\Delta_\omega}-\frac{m}{\Delta_\omega}\tan\left(-\frac{m}{2}t+\tan^{-1}M\right)\right),\\
    M &=  \left(\frac{K'-\Delta_\omega\tan{(\theta_{i}-\Psi(0))}}{m}\right),
\end{align}
where $\Delta_\omega = \omega_{i}-\omega_0,K'=K{R_c}$, {and} $m=\sqrt{\Delta_\omega^2-K'^2}$.

\begin{figure}
    \centering
    \includegraphics[width=1.0\linewidth]{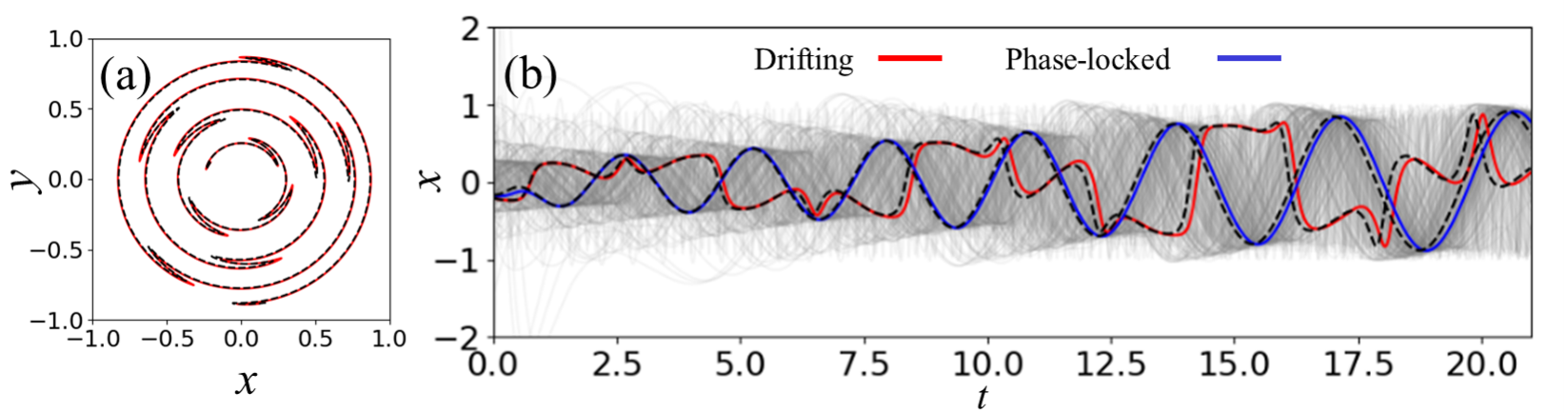}
    \caption{Dynamics of globally coupled KSL oscillators. (a) Trajectories in the $(x, y)$ plane of a single drifting oscillator. (b) Temporal evolution of the $x$ variables of the oscillators.
    In both panels, the red line represents the dynamics of a single drifting oscillator obtained numerically, and the black dotted line represents the analytical solution. In (b), the gray lines represent the $x$ variables of $50$ other oscillators.}
    \label{fig:multi-body}
\end{figure}

Figure~\ref{fig:multi-body}~(a) and (b) illustrate the results of numerical simulations using $N = 2000$ KSL oscillators.
The phase $\theta$ and amplitude $r$ of each oscillator was sampled from Eq.~\eqref{eq:theta_dis} and from a uniform distribution on the interval $[-2,\,0.5]$, respectively.
The parameters are set to $K=5$, $a=c=0.1$, $d=1$, $\Delta = 1$, and $\omega_0=\pi/2$.
In panel (a), the red line represents the trajectory of one drifting oscillator in the $(x, y)$ plane, while the black dashed line shows its corresponding analytical solution, demonstrating complex but analytically solvable trajectories. 
The change in the direction of rotation originates from the alternation between states with and without phase slips.

In panel (b), the same results for the drifting oscillator in panel (a), 
one drift-locked ocillator (blue) and randomly selected $50$ other oscillators (gray)
are plotted on the $t-x$ plane, showing irregular dynamics. 
Note that the whole system is in a steady synchronized state in terms of phase dynamics, even though the amplitudes of individual oscillators exhibit irregular evolution toward the underlying limit cycle.\\
\par

\begin{figure}[t]
    \centering
    \includegraphics[width=1.0\linewidth]{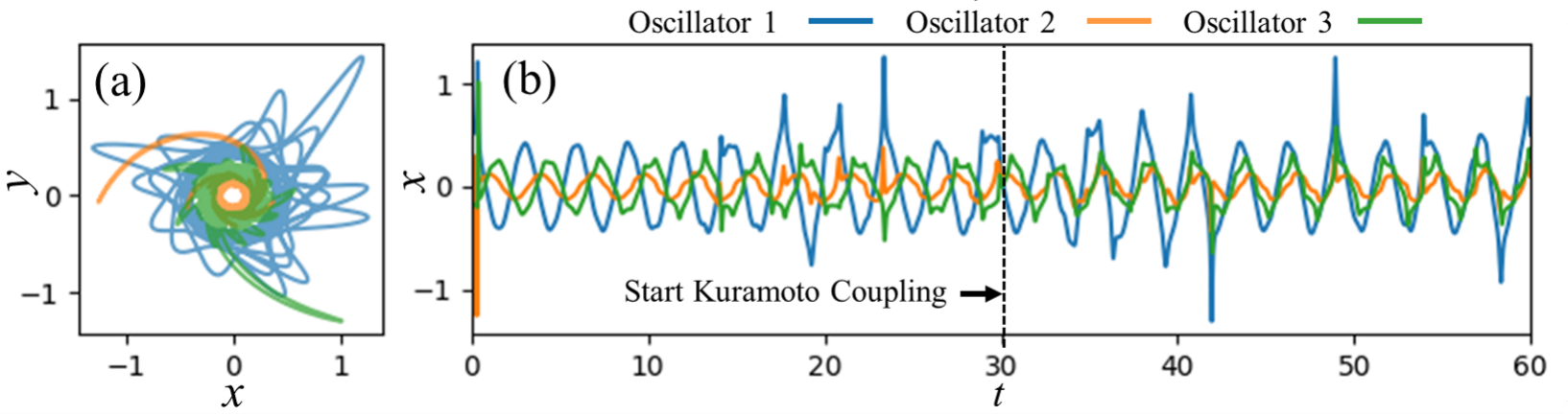}
    \caption{Dynamics of 
    the KSL-Lorenz model.
    (a) Trajectories in the $(x, y)$ plane of three oscillators.
    (b) Temporal evolution of the $x$ variables of the oscillators.
    At time $t=30$, the Kuramoto phase coupling is activated.}
    \label{fig:KR}
\end{figure}

\textbf{KSL-Lorenz model}.
Finally, as an illustrative example of complex non-decaying amplitude dynamics, we construct a coupled-oscillator model in which the phase dynamics is described by the Kuramoto model, whereas the amplitude dynamics is governed by the chaotic Lorenz system~\cite{lorenz2017deterministic}.
Specifically, we choose the coefficients $G_{i}$ 
in the coupling function Eq.~\eqref{eq:coupling0} as
$
    {G}_{1}(\bm{x}_{1},\bm{x}_{2},\bm{x}_{3}) = \sigma\Gamma(\bm{x}_{2}),
    {G}_{2}(\bm{x}_{1},\bm{x}_{2},\bm{x}_{3}) = \Gamma(\bm{x}_{1})(\alpha-\Gamma(\bm{x}_{3})),
    {G}_{3}(\bm{x}_{1},\bm{x}_{2},\bm{x}_{3}) = \Gamma(\bm{x}_{1})\Gamma(\bm{x}_{2}).
$
We can then confirm that the dynamics of the oscillator amplitudes $r_{1,2,3} = \Gamma({\bm x}_{1,2,3})$ obey the Lorenz vector field 
$
    \dot{r_1} = \sigma (r_2 - r_1), 
    \dot{r_2} = r_1(\alpha - r_3) - r_2, 
    \dot{r_3} = r_1r_2 - \beta r_3.
$
Here, the linear damping terms arise from the spontaneous decay of the amplitude variables, and we choose the parameters (Floquet exponents) $a_i = \lambda_i$ of the SL oscillators as $a_1=\sigma/2$, $a_2=1/2$, and $a_3=\beta/2$.

Figure~\ref{fig:KR} shows the dynamics of the three SL variables exhibiting Kuramoto-Lorenz dynamics, where (a) shows the dynamics of the SL variables on the $(x, y)$ plane and (b) plots the time evolution of the $x$ variables. 
The parameters of the Lorenz model are chosen as $\sigma=10$, $\beta=8/3$, and $\alpha=28$,
and the Kuramoto coupling with strength $K=0.2$ is introduced to the phase variables after $t=30$.
The initial phases are set to $0$, $2\pi/3$, and $4\pi/3$, respectively.
Since the amplitude variables of the SL oscillators defined in Eq.~(\ref{eq:ampfunction}) and bounded as $r_i\le c_i$, constant offsets $(-16,-28,-48)$ were added to the Lorenz variables 
$(r_1, r_2, r_3)$ throughout the simulations so that the resulting amplitude trajectories remain within the admissible range.
Despite the complex amplitude dynamics, the timing of the peaks becomes synchronized after $t=30$, indicating that phase synchronization is established by the Kuramoto coupling.
\\
\par

\textbf{Discussion.}

In summary, we have presented a framework for constructing systems of coupled oscillators that preserve Kuramoto phase dynamics globally in the state space, regardless of amplitude dynamics, based on phase-amplitude coordinates.
As a solvable example, we proposed the KSL model, which exhibits intriguing phase-synchronized dynamics away from the limit cycle, as well as the KSL-Lorenz model, which demonstrates phase synchronization despite chaotic amplitude dynamics.
While we focused on KSL oscillators with Kuramoto coupling for analytical tractability, our method is generally applicable to other limit-cycle oscillators and arbitrary coupling topologies~\cite{arenas2008synchronization,battiston2020networks,bick2018chaos,rodrigues2016kuramoto}.
The phase and amplitude functions can be evaluated numerically for other models or estimated from time-series data~\cite{namura2022estimating,yawata2024phase}.
The coupling can be arbitrary functions of all oscillator states, including higher-order interactions~\cite{millan2020explosive,bick2024higher,leon2024higher}.
Although not addressed in this study, external inputs can also be incorporated similarly to control global phase dynamics.
Thus, the proposed framework extends the notion of synchronization beyond the conventional weak-coupling paradigm, enabling the description of synchronization phenomena with strong amplitude fluctuations.

%\bibliography{main}% Produces the bibliography via BibTeX.

\end{document}